\documentclass{research}
\usepackage{color}
\usepackage{epsfig,amssymb,amsmath,amsthm,amsfonts,amsbsy,mathrsfs}
\usepackage{graphics}
\usepackage{graphicx,color}
\usepackage{amsmath}
\usepackage{amssymb}
\usepackage[version=3]{mhchem} 
\usepackage{float}
\usepackage{booktabs}
\usepackage{longtable}
\usepackage{siunitx}  
\usepackage{caption}   
\usepackage{geometry}

\title{High-throughput Screening of Ferrimagnetic Semiconductors With Ultrahigh N$\acute{e}$el Temperature}

\author{Haidi Wang$^{1}$, Qingqing Feng$^{2}$, Shuo Li$^{2}$, Wei Lin$^{1}$, Weiduo Zhu$^{1}$, Zhao Chen$^{1}$, Zhongjun Li$^{1}$, Xiaofeng Liu$^{1*}$ \& Xingxing Li$^{2*}$ }

\begin{document}
	
	\maketitle
	
	\begin{affiliations}
		\item School of Physics, Hefei University of Technology, Hefei, Anhui 230601, China
		\item Hefei National Laboratory for Physical Sciences at Microscale, Department of Chemical Physics, and Synergetic Innovation Center of Quantum Information and Quantum Physics, University of Science and Technology of China, Hefei, Anhui 230026, China
	\end{affiliations}

	\begin{addendum}
		
		\item[Correspondence]  Xiaofeng Liu (lxf@hfut.edu.cn) 
		\& Xingxing Li (lixx@ustc.edu.cn)  \\

	\end{addendum}

\begin{abstract}
Ferrimagnetic semiconductors, integrated with net magnetization, antiferromagnetic coupling and  semi-conductivity, have constructed an ideal platform for spintronics. For practical applications, achieving high N$\acute{e}$el temperatures ($T_{\mathrm{N}}$) is very desirable, but remains a significant challenge. Here, via high-throughput density-functional-theory calculations, we identify 19 intrinsic ferrimagnetic semiconductor candidates from nearly 44,000 structures in the Materials Project database, including 10 ferrimagnetic bipolar magnetic semiconductors (BMS) and 9 ferrimagnetic half semiconductors (HSC). Notably, the BMS \ce{NaFe5O8} possesses a high $T_{\mathrm{N}}$ of 768 K. By element substitutions, we obtain an HSC \ce{NaFe5S8} with  a $T_{\mathrm{N}}$ of 957 K and a BMS \ce{LiFe5O8} with a $T_{\mathrm{N}}$ reaching 1059 K. Our results pave a promising avenue toward the development of ferrimagnetic spintronics at ambient temperature.
\end{abstract}
\section*{Introduction}

Spintronics has attracted great attention because it aims to utilize the intrinsic spin freedom of electrons to develop high-performance electronic devices.\cite{5388780,RevModPhys.76.323,RevModPhys.80.1517,Hirohata2020} In contrast to conventional electronics, which relies primarily on electron charge freedom, spintronics takes advantage of both the charge and spin properties of electrons, enabling the exploration of new functionalities. Spintronics offers several advantages over traditional electronics, including reduced energy consumption, improved operation speed, increased information storage capacity, and the potential for non-volatility of information.\cite{Wolf2001} However, its development faces three major challenges: the generation and injection of spins, long-range spin transport, and the manipulation and detection of spins.\cite{Awschalom2007,Dieny2020} 

To address these issues, solutions rely on advancements in device fabrication technologies and processes on the one hand, and on the other hand, on finding suitable spintronics materials, such as half metals (HM),\cite{PhysRevLett.50.2024,PhysRevLett.74.1171} half semiconductors (HSC),\cite{Li2016NSR, PhysRevB.67.180401} and bipolar magnetic semiconductors (BMS) et al.\cite{Li2012,Li2013h,Yuan2013,chen2024,li2022} Specifically, HM materials conduct electrons of one spin orientation while acting as insulators or semiconductors for the opposite spin, enabling 100\% spin-polarized electron generation, injection, and transport. 
HSC are unique in that they act as semiconductors in one spin channel and insulators in the other, allowing them to generate 100\% spin-polarized electrons and holes. BMS, with valence and conduction band edges spin-polarized in different direction, allow the electrical manipulation of carrier spin orientation, which is critical for developing high-performance spintronics devices. Notably, employing the electric field to manipulate spin polarization, as opposed to the magnetic field, gives considerable benefits due to the ease of local application and precise control of electric fields.  

These above-mentioned HM, HSC, and BMS materials offer essential functionalities such as spin injection, spin transport, magnetoelectric coupling, and spin torque sensing through the manipulation of magnetic properties.\cite{Li2016e} However, two key challenges must be addressed before they can be broadly utilized practically. First, many of these materials exhibit low magnetic transition temperatures,\cite{Li2014n} limiting their practical application at room temperature. Second, most magnetic materials are synthesized through doping\cite{Zhang2021} or chemical modification,\cite{Li2012} resulting in a scarcity of intrinsic magnetic semiconductors. Consequently, extensive research leveraging modern material design strategies is crucial to discover intrinsic semiconductor materials with room-temperature magnetism.

Recently, high-throughput computational material screening has become widely utilized in material design.\cite{NIST2011,Urban2016,Yan2017,Chen2019,Horton2019,zhang2023magnetic,liu2018screening,research_wang} This strategy offers a significant advantage by enabling efficient screening of materials with specific properties, thereby accelerating the discovery and optimization process. Notable studies include the high-throughput computational screening of magnetic electrides, which have potential applications in spintronics, topological electronics, electron emission, and high-performance catalysts.\cite{zhang2023magnetic} Another study involved high-throughput first-principles simulations to screen 2D materials that exhibit a non-magnetic to ferromagnetic (FM) transition upon hole doping, identifying 122 materials displaying hole-doping-induced ferromagnetism.\cite{liu2018screening} Additionally, a high-throughput screening for BMS materials has been proposed in our previous work and led to the discovery of  a serials of BMS with FM ordering.\cite{research_wang} Deng et al. identified 11 promising BMS candidates screened from 2DMatPedia database.\cite{acsaelm.2c00464} These studies exemplify the effectiveness of high-throughput computational material screening in accelerating the exploration of new spintronics materials.

In this study, we conducted high-throughput density-functional-theory calculations for screening of ferrimagnetic semiconductors from the Materials Project database (MPD),\cite{Jain2013} built upon previous BMS screening strategies.\cite{research_wang} Ferrimagnetic semiconductors were specifically targeted due to their generally strong antiferromagnetic coupling compared to FM semiconductors, which are often associated with elevated magnetic transition temperatures and non-zero net magnetic moments.\cite{kim2022ferrimagnetic} We identified 10 ferrimagnetic BMS materials and 9 ferrimagnetic HSC materials, providing guidelines for the development spintronics based on ferrimagnetic semiconductors.

\section{Results}

\begin{figure*}[!hbt]
	\centering
	\includegraphics[width=0.8\textwidth]{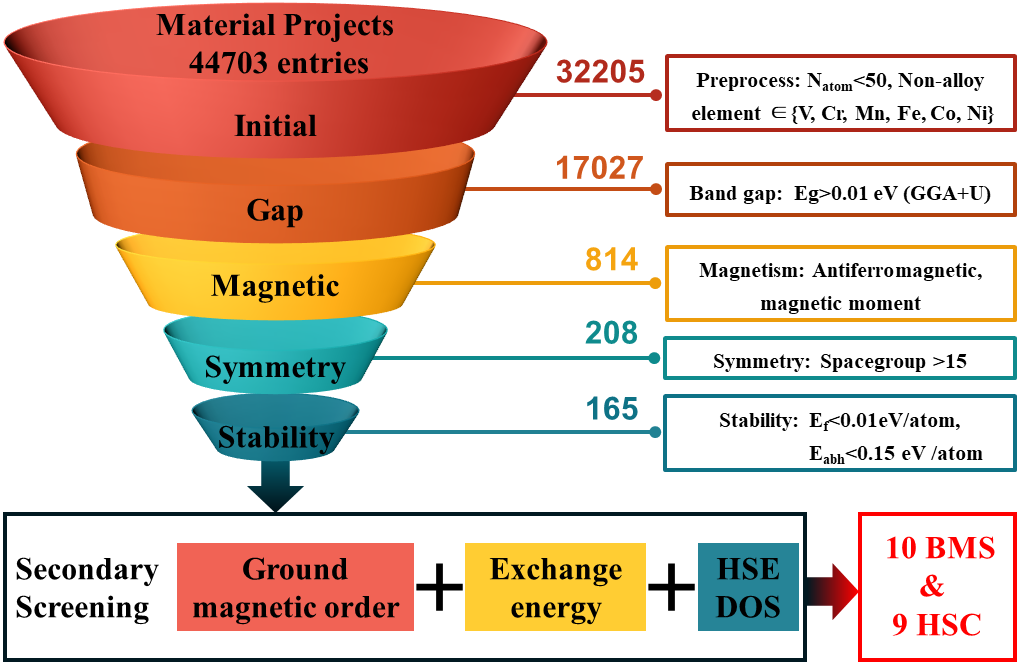}
	\caption{Schematic representation of the material selection protocol utilizing primary and secondary screening phases for ferrimagnetic semiconductors. The primary screening encompasses an initial, gap, magnetism, symmetry, and stability filter, systematically narrowing down the initial 44,703 entries from the Material Projects database. The secondary screening further refines the candidates through assessments of ground magnetic order, exchange energy, and density of states at the HSE06 level, culminating in 10 BMS and 9 HSC candidates.}
	\label{fig:1}
\end{figure*}

In the Figure \ref{fig:1}, we outline a rigorous two-stage high-throughput screening protocol used to identify magnetic semiconductor materials, which has been successfully applied for the screening of BMS materials.\cite{research_wang} The initial screening comprises an initial filter, gap, magnetism, symmetry, and stability filter, sequentially applied to the initial 44,703 inorganic crystal structures obtained from the MPD.

\begin{table*}[!h]
	\centering
	\caption{\label{Table:1} Properties of obtained ferrimagnetic semiconductors: Materials Project ID (MP-ID), formula,  space group symmetry, formation energy $E_{\mathrm{f}}$ (eV/atom), energy above hull $E_{\mathrm{abh}}$ (eV/atom), energy difference ($\Delta$$E$) between ferromagnetic and antiferromagnetic configuration per magnetic atom (eV/atom), magnetic moment per formula unit cell $M$ ($\mu_{\mathrm{B}}$/f.u.). The spin-flip gap in valence band $\Delta_{\mathrm{V}}$ (eV) and conduction band $\Delta_{\mathrm{C}}$ (eV), and the band gap $E_{\mathrm{g}}$ (eV). The classification of ferrimagnetic semiconductors including half semiconductors (HSC) and bipolar magnetic semiconductors (BMS) in PBE and HSE level.}
	\resizebox{0.8\textwidth}{!}{%
		\setlength{\tabcolsep}{0.7mm}{
			\begin{tabular}{lcccccccccc}
				\hline
				MP-ID & Formula & Symmetry & $E_{\mathrm{f}}$ & $E_{\mathrm{abh}}$ & $\Delta$$E$ & $M$ & $\Delta_{\mathrm{V}}$ & $\Delta_{\mathrm{C}}$ & $E_{\mathrm{g}}$ & classfication (PBE/HSE) \\
				\hline
				mp-759974  & \ce{NaFe5O8}     & $R\overline{3}m$   & -1.692 & 0.039 & -0.216 & 5  & 1.669 & 0.432 & 0.462 & BMS/BMS \\
				mp-35596   & \ce{Fe2NiO4}     & $Imma$             & -1.589 & 0.000 & -0.179 & 4  & 1.837 & 0.513 & 0.347 & BMS/BMS \\
				mp-1214367 & \ce{BaTbFe4O7}   & $P6_{3}mc$         & -2.163 & 0.080 & -0.142 & 2  & 1.134 & 0.238 & 0.224 & BMS/HSC \\
				mp-39239   & \ce{SrLaMnRuO6}  & $R3$               & -2.598 & 0.094 & -0.133 & 2  & 0.329 & 0.244 & 1.960 & HSC/HSC \\
				mp-1221978 & \ce{MgZn(FeO2)4} & $R3m$              & -1.914 & 0.039 & -0.081 & 0  & 1.992 & 0.015 & 0.015 & BMS/-\\
				mp-674482  & \ce{MnFeO3}      & $Ibca$             & -1.846 & 0.012 & -0.062 & 32 & 0.547 & 0.624 & 1.202 & HSC/HSC \\
				mp-755278  & \ce{Li2VCrO4}    & $I\overline{4}m2$  & -2.413 & 0.036 & -0.060 & 0  & 1.611 & 0.078 & 0.133 & BMS/BMS \\
				mp-1035281 & \ce{Mg14CoNiO16} & $Pmmm$             & -2.807 & 0.008 & -0.058 & 0  & 2.445 & 0.050 & 0.201 & BMS/BMS \\
				mp-753261  & \ce{Li5MnCr3O8}  & $R\overline{3}m$   & -2.214 & 0.087 & -0.048 & 4  & 1.526 & 1.477 & 0.110 & BMS/BMS \\
				mp-760089  & \ce{LiMn2OF3}    & $Pnma$             & -2.661 & 0.096 & -0.040 & 0  & 1.473 & 1.259 & 0.353 & BMS/BMS \\
				mp-1076580 & \ce{La5Sm3Cr6(FeO12)2} & $Amm2$       & -2.972 & 0.127 & -0.034 & 11 & 2.209 & 0.228 & 0.067 & BMS/-   \\
				mp-1639830 & \ce{Li2Fe(CoO3)2}& $Aea2$             & -1.567 & 0.100 & -0.032 & 2  & 0.761 & 0.179 & 0.179 & HSC/HSC \\
				mp-18759   & \ce{Mn3O4}       & $I4_{1}/amd$       & -2.048 & 0.000 & -0.029 & 0  & 1.154 & 0.049 & 0.049 & BMS/BMS \\
				mp-1571056 & \ce{Li2Mn2FeO6}  & $C222_{1}$         & -1.948 & 0.086 & -0.026 & 4  & 0.890 & 0.846 & 1.618 & HSC/HSC \\
				mp-1573004 & \ce{Li2Mn2FeO6}  & $Aea2$             & -1.956 & 0.078 & -0.026 & 4  & 1.014 & 0.754 & 1.631 & HSC/-\\
				mp-759988  & \ce{Li3MnFe3O8}  & $P6_{3}mc$         & -1.898 & 0.059 & -0.024 & 24 & 1.320 & 0.404 & 1.913 & BMS/BMS \\
				mp-1663998 & \ce{Li2Mn3FeO8}  & $Cmc2_{1}$         & -1.977 & 0.032 & -0.023 & 10 & 0.336 & 1.207 & 1.314 & HSC/HSC \\
				mp-1189831 & \ce{Gd3FeO6}     & $Cmc2_{1}$         & -3.340 & 0.006 & -0.021 & 4  & 2.780 & 0.041 & 0.436 & BMS/BMS \\
				mp-769631  & \ce{Li2Mn3NiO8}  & $R\overline{3}m$   & -1.936 & 0.012 & -0.020 & 7  & 0.735 & 0.889 & 1.145 & BMS/BMS \\
				mp-1275319 & \ce{Li2MnCrO4}   & $C222_{1}$         & -2.260 & 0.022 & -0.017 & 7  & 1.191 & 0.843 & 0.789 & HSC/HSC \\
				mp-1290535 & \ce{Li2Fe2(CO3)3}& $Imm2$             & -1.843 & 0.116 & -0.016 & 7  & 3.423 & 0.030 & 0.207 & HSC/- \\
				mp-1229097 & \ce{AlCrFeO4}    & $Imma$             & -2.540 & 0.000 & -0.015 & 2  & 2.425 & 0.533 & 0.449 & HSC/HSC \\
				mp-1344634 & \ce{CaNiAsO5}    & $P2_{1}2_{1}2_{1}$ & -1.978 & 0.054 & -0.011 & 3  & 0.114 & 0.227 & 2.271 & BMS/HSC \\
				\hline
			\end{tabular}
		}
	}
\end{table*}

The \textbf{Initial Filter} serves as the basis for screening materials that include magnetic atoms {V, Cr, Mn, Fe, Co, or Ni}. We exclude entirely metal-composed alloy materials and structures with more than 50 atoms per unit cell due to their metallic behavior and consideration of computational resources, resulting in 32,205 entries. The \textbf{Gap Filter} is then used to select potential semiconductors. It should be noted that the GGA+U method might underestimate band gaps. Therefore, we adopted a relatively relaxed standard of $E_{\mathrm{g}}$ $>$ 0.01 eV to ensure the materials with the semiconductor characteristics, narrowing the selection to 17,027 entries. Subsequently, the \textbf{Magnetism Filter} assesses the magnetic properties, selecting materials with antiferromagnetic order and net magnetic moments, yielding 814 potential entries. 
The \textbf{Symmetry Filter} then imposes further restrictions based on crystal symmetry with space group number greater than 15 (resulting in 208 structures), which can enhance computational efficiency, simplify theoretical model, ensure the accuracy of simulations, and facilitate in-depth studies of electronic and magnetic properties, such as excahange interaction and N$\acute{e}$el temperature. Lastly, the \textbf{Stability Filter} evaluates the thermodynamic stability with formation energy ($E_{\mathrm{f}}$ $<$ 0.01 eV/atom) and energy above hull ($E_{\mathrm{abh}}$ $<$ 0.15 eV/atom) in SCAN level by using the Materials RESTful API.\cite{Wang2020, Ong_2015}

In the second screening stage, a detailed evaluation of magnetic properties is performed, including the ground magnetic order, exchange energy, and density of states (DOS) at the HSE06 level. The ground state structure is determined by comparing the energies of the ferromagnetic state and antiferromagnetic states. We enumerate all possible magnetic configurations for unit cells with six or fewer magnetic atoms. In the case of more than six magnetic atoms, we determine the magnetic ground state by constructing supercells ($2 \times1 \times 1$, $1 \times 2 \times 1$, and $1 \times 1 \times 2$), which has been employed in previous work.\cite{research_wang} To avoid redundant calculations, we use XtalComp\cite{LONIE2012690} to exclude duplicate structures. Furthermore, we systematically distinguish and treat them as either high-spin or low-spin states for these structures containing Co elements.\cite{Kremer1982} Then, the energy ($\Delta E_{\mathrm{AFM-FM}}$) of the antiferromagnetic configurations averaged over the magnetic atoms is calculated relative to the energy of the ferromagnetic configuration. The large $\Delta E_{\mathrm{AFM-FM}}$
 is a hallmark of strong antiferromagnetic coupling, which results in high N$\acute{e}$el temperature ($T_{\mathrm{N}}$).

\begin{figure*}[!hbt]
	\centering
	\includegraphics[width=0.95\textwidth]{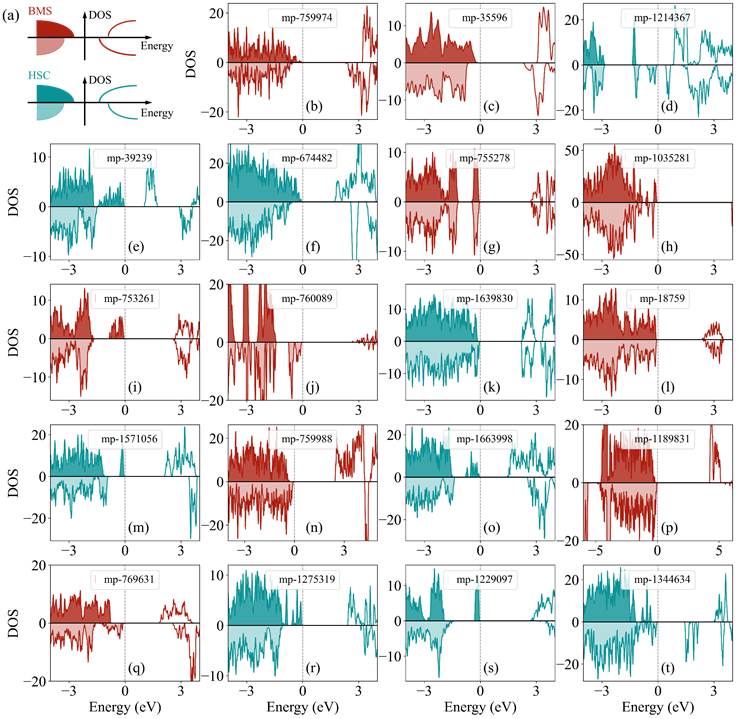}
	\caption{(a) The DOS sketch map of BMS and HSC. (b)-(t)  HSE06 level DOSs for \ce{NaFe5O8}(mp-759974), \ce{Fe2NiO4}(mp-35596), \ce{BaTbFe4O7}(mp-1214367), \ce{SrLaMnRuO6}(mp-39239), \ce{MnFeO3}(mp-674482), \ce{Li2VCrO4}(mp-755278), \ce{Mg14CoNiO16}(mp-1035281), \ce{Li5MnCr3O8}(mp-753261), \ce{LiMn2OF3}(mp-760089), \ce{Li2Fe(CoO3)2}(mp-1639830), \ce{Mn3O4}(mp-18759), \ce{Li2Mn2FeO6}(mp-1571056), \ce{Li3MnFe3O8}(mp-759988), \ce{Li2Mn3FeO8}(mp-1663998), \ce{Gd3FeO6}(mp-1189831), \ce{Li2Mn3NiO8}(mp-769631), \ce{Li2MnCrO4}(mp-1275319), \ce{AlCrFeO4}(mp-1229097), \ce{CaNiAsO5}(mp-1344634), respectively. }
	\label{fig:2}
\end{figure*}

\begin{figure*}[!hbt]
	\centering
	\includegraphics[width=0.95\textwidth]{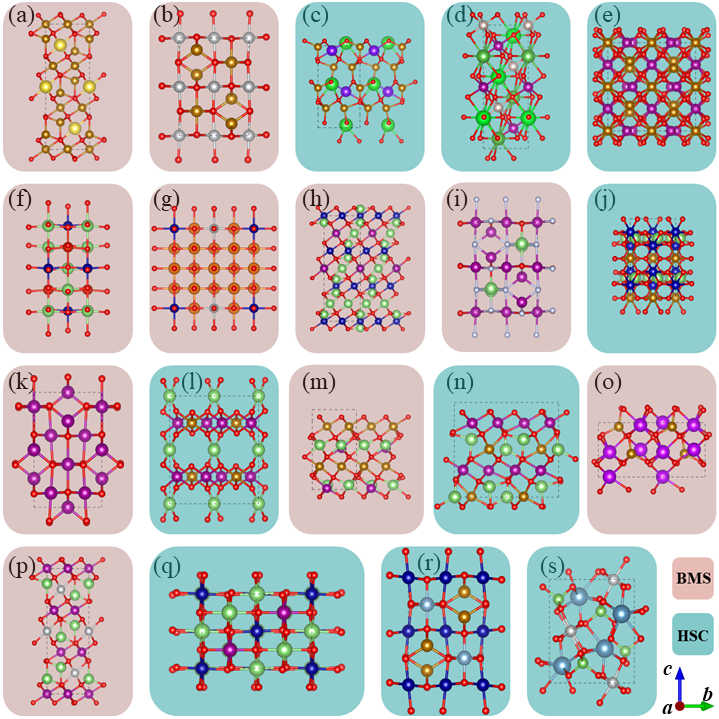}
	\caption{(a)-(s) Geometry structures of \ce{NaFe5O8}(mp-759974), \ce{Fe2NiO4}(mp-35596), \ce{BaTbFe4O7}(mp-1214367), \ce{SrLaMnRuO6}(mp-39239), \ce{MnFeO3}(mp-674482), \ce{Li2VCrO4}(mp-755278), \ce{Mg14CoNiO16}(mp-1035281), \ce{Li5MnCr3O8}(mp-753261), \ce{LiMn2OF3}(mp-760089), \ce{Li2Fe(CoO3)2}(mp-1639830), \ce{Mn3O4}(mp-18759), \ce{Li2Mn2FeO6}(mp-1571056), \ce{Li3MnFe3O8}(mp-759988), \ce{Li2Mn3FeO8}(mp-1663998), \ce{Gd3FeO6}(mp-1189831), \ce{Li2Mn3NiO8}(mp-769631), \ce{Li2MnCrO4}(mp-1275319), \ce{AlCrFeO4}(mp-1229097), \ce{CaNiAsO5}(mp-1344634),   respectively}
	\label{fig:3}
\end{figure*}

Following the screening scheme, we have identified 23 ferrimagnetic semiconductors based on exchange interaction rankings validated using the HSE06 functional, leading to the identification of ferrimagnetic semiconductor candidates, as shown in Table \ref{Table:1} and S1. They can be categorized into 10 BMS and 9 HSC, as demonstrated in Figure~\ref{fig:2} and ~\ref{fig:3}. The electronic structures calculated based on the PBE functional are consistent with the HSE06 level for most screened candidates. For \ce{BaTbFe4O7}(mp-1214367), it belongs to BMS at PBE functional while classified as an HSC at HSE06 functional. It is also noted that the valence band edges calculated with HSE06 functional degenerate for spin-up and spin-down polarized bands in these case of \ce{MgZn(FeO2)4}(mp-1221978), \ce{La5Sm3Cr6(FeO12)2}(mp-1076580), \ce{Li2Mn2FeO6}(mp-1573004), and \ce{Li2Fe2(CO3)3}(mp-1290535). 

Based on the HSE06 functional, we analyzed the density of states (DOS) for both HSC and BMS materials, as shown in Figure~\ref{fig:2}(a). In HSC materials, the valence band maximum (VBM) and conduction band minimum (CBM) are fully spin-polarized in the same spin channel, which can generate 100\% spin-polarized electrons and holes through thermal, optical excitation, and electrical gating.\cite{Li2016NSR} BMS can create completely spin-polarized currents in a reversible spin-polarization direction by applying a gate voltage. When a negative gate voltage is applied, the Fermi level shifts into the spin-flip gap in VB, leading to half-metallic conduction with carriers fully spin-polarized in the spin-up direction. Conversely, under a positive gate voltage, the Fermi level shifts into the spin-flip gap in CB, resulting in carriers being fully spin-polarized in the spin-down direction. The results of this analysis are presented in Figure \ref{fig:2}(b)-(t), which illustrates that the 10 BMS materials and 9 HSC materials studied all exhibit the aforementioned spin-polarization characteristics. Additionally, our statistical analysis revealed that among these BMS and HSC candidates, 17 contain alkali and alkaline earth metals, constituting 89\% of the total (Figure \ref{fig:3}). For the 11 BMS materials mentioned in previous work,\cite{research_wang} 73\% contain alkali metals. Alkali metals are known to donate electrons to maintain electrical neutrality, enabling non-metal elements to achieve a full outer electron shell, thereby conferring semiconductor properties. Bimetallic compounds offer a potential solution to this configuration, particularly in ferromagnetic structures. These findings provide valuable insights for the future design of BMS and HSC materials.

\begin{figure}[!hbt]
	\centering
	\includegraphics[width=1.0\textwidth]{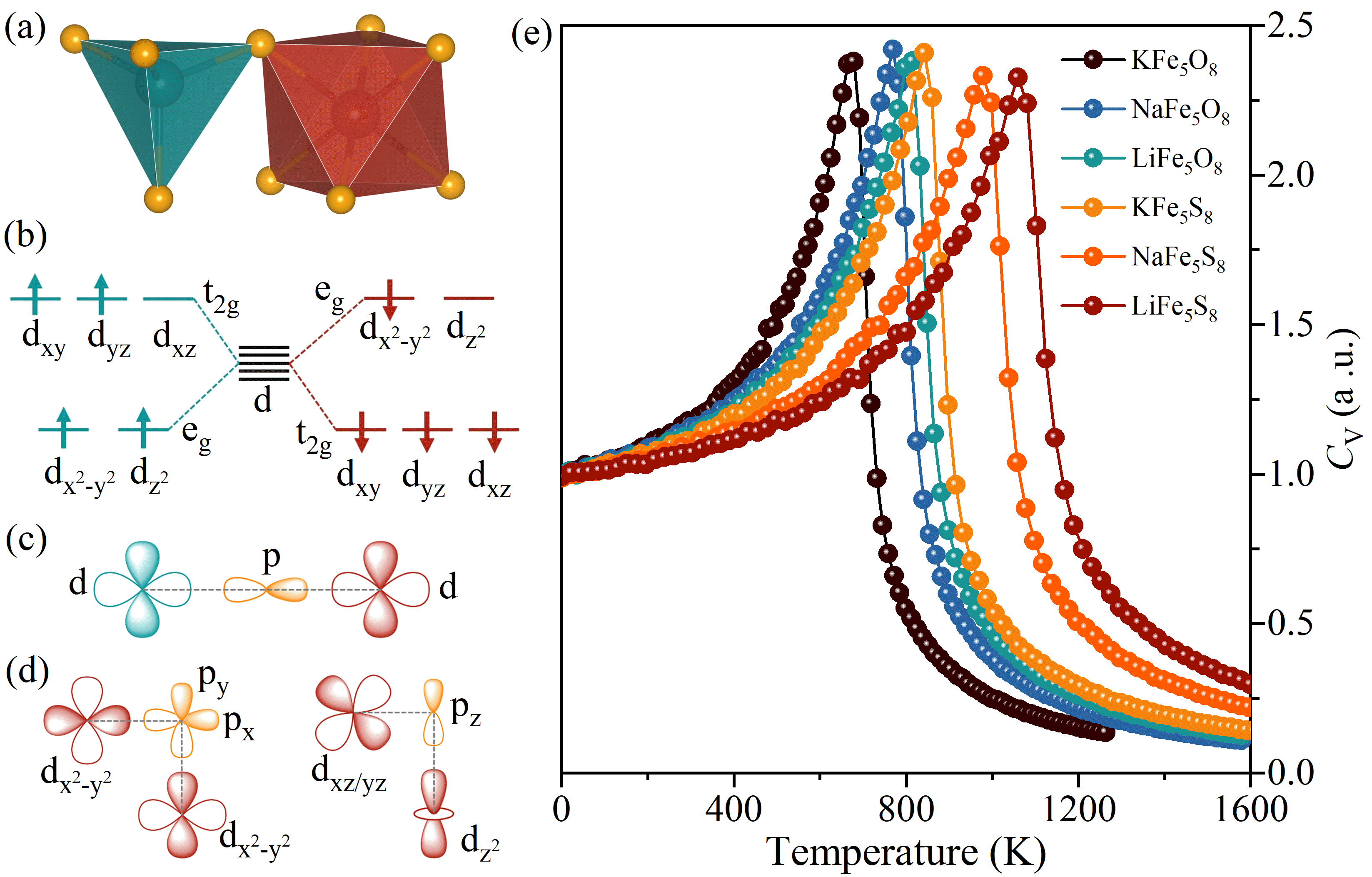}
	\caption{(a) Geometry of four-coordinated (blue) and six-coordinated (red) Fe atoms in the \ce{NaFe5O8}. (b) Splitting of the d orbital in the tetrahedral and octahedral crystal fields. (c) AFM superexchange interaction through the 180$^{\circ}$ d-p-d path. (d) Two FM superexchange paths: ferromagnetic superexchange interaction between the occupied $d_{\mathrm{x^2-y^2}}$ orbitals through the $p_{\mathrm{x/y}}$ orbitals of O atoms; FM superexchange path between the occupied $d_{\mathrm{xz/yz}}$ and empty $d_{\mathrm{z^2}}$ orbitals through the $p_{\mathrm{z}}$ orbital of O atoms. (e) $C_{\mathrm{V}}$ to temperature for \ce{MFe5X8} (M=\{Li, Na, K\}, X=\{O, S\}) materials through the Monte Carlo simulation based on the Heisenberg model.}
	\label{fig:4}  
\end{figure}

Particularly, materials \ce{NaFe5O8}(mp-759974), \ce{Fe2NiO4}(mp-35596), \ce{BaTbFe4O7}(mp-1214367), \ce{SrLaMnRuO6}(mp-39239), with $\Delta E_{\mathrm{AFM-FM}}$ exceeding 100 meV/atom, are predicted to possess high $T_{\mathrm{N}}$. Substitution of elements within the same group is a commonly employed strategy to modulate chemical properties and find the new materials effectively.\cite{10.1063/5.0070846,Haastrup2018,Zhou2019} Herein, we investigate the substitution of alkali and chalcogen elements within the \ce{NaFe5O8} structure, aiming to explore the potential for obtaining enhanced magnetic semiconductor materials. As is show in the Table \ref{table:2}, the energy, electronic and magnetic properties of \ce{MFe5X8} (M=\{Li, Na, K\}, X=\{O, S\}) materials are presented. It can be found that in the \ce{MFe5X8} family (see Figure \ref{fig:4}(a)), the iron (Fe) atoms are coordinated with four and six oxygen (O) atoms in tetrahedral and octahedral crystal fields, respectively. Six-coordinated Fe atoms form a 2D layered structure and are alternately arranged with four-coordinated iron atoms along the direction perpendicular to the plane. According to the crystal field theory, the d orbital are split to two groups in the tetrahedral and octahedral crystal fields: $t_{\mathrm{2g}}$ ($d_{\mathrm{xy}}$, $d_{\mathrm{xz}}$, and $d_{\mathrm{yz}}$) and $e_{\mathrm{g}}$ ($d_{\mathrm{x^2-y^2}}$ and $d_{\mathrm{z^2}}$). In the octahedral crystal field, the energy of $t_{\mathrm{2g}}$ is lower than $e_{\mathrm{g}}$, while the opposite is true in the tetrahedral field, as shown in Figure~\ref{fig:4}(b). Based on our calculations, the magnetic moments of two types of Fe are 4 $\mu_{\mathrm{B}}$. In the case of four-coordinated Fe atoms, its $d^{4}$ states are $e_{\mathrm{g}}^2 t_{\mathrm{2g}}^2$, while for the eight-coordinated Fe, the $t_{\mathrm{2g}}$ states and one of $e_{\mathrm{g}}$ states are filled to form a $t_{\mathrm{2g}}^3 e_{\mathrm{g}}^1$ state. Since the angle Fe-O-Fe between the four and eight-coordinated Fe is 124$^{\circ}$, the magnetic exchange interaction prefers to be antiferromagnetic (Figure~\ref{fig:4}c) according to the GKA rules (Figure~\ref{fig:4}c).\cite{goodenough1955theory,anderson1959new,kanamori1960crystal} For the magnetic moment of eight-coordinated Fe, the occupied $d_{\mathrm{x^2-y^2}}$ can couple with each other in parallel through the interaction with $p_{\mathrm{x}}$ orbital of O atoms. The occupied $d_{\mathrm{xz/yz}}$ can also couple with the unoccupied $d_{\mathrm{z^2}}$ through virtual hopping.\cite{jacs.8b07879} In all, the magnetic moments of six-coordinated Fe in the 2D plane are arranged ferromagnetically through the d-p-d superexchange interaction (Figure \ref{fig:4}d), since two Fe atoms are connected through sharing O atoms with nearly 90$^{\circ}$ Fe-O-Fe. Therefore, magnetic moments of the six-coordinated Fe atoms are coupled together in parallel and anti-parallel with that of four-coordinated Fe atoms, leading to a ferrimagnetic configuration. Finally, we use the Property Analysis and Simulation Package (PASP)\cite{10.1063/5.0043703} along with the Monte Carlo simulation method,\cite{PhysRevB.73.214412} based on the classical Heisenberg Hamiltonian, to estimate the N$\acute{e}$el 
temperature:

\begin{equation}
	\label{eq2}
	H=-\sum_{i,j}{J_{i,j}S_{i}S_{j}}  
\end{equation}
where $J_{i,j}$ is the exchange parameter and $S$ is the spin of magnetic atoms. The calculated results show that all obtained 4 BMSs and 2 HSCs with high $T_{\mathrm{N}}$, especially the $T_{\mathrm{N}}$ of \ce{LiFe5S8} is up to 1059.3 K.

\begin{table*}[!ht]
	\centering
	\caption{Properties of \ce{MFe5X8} (M=\{Li, Na, K\}, X=\{O, S\}) materials.}
	\label{table:2}
	\resizebox{0.8\textwidth}{!}{ 
	\begin{tabular}{ccccccc}
		\hline
		Properties & \ce{NaFe5O8} & \ce{NaFe5S8} & \ce{LiFe5O8} & \ce{LiFe5S8} & \ce{KFe5O8} & \ce{KFe5S8} \\
		\hline
		E(FM/eV)             & 3.326 & 4.520 & 3.420 & 4.782  & 3.091 & 4.212 \\
		E(FIM-1/eV)          & 0     & 0     & 0     & 0      & 0     & 0     \\
		E(FIM-2/eV)          & 2.798 & 3.610 & 2.785 & 3.593  & 2.757 & 3.605 \\
		E(FIM-3/eV)          & 1.620 & 2.091 & 1.667 & 2.226  & 1.494 & 1.874 \\
		E(FIM-4/eV)          & 2.124 & 2.649 & 2.170 & 2.776  & 2.002 & 2.481 \\
		$T_{\mathrm{N}}$ (K) & 768.8 & 957.9 & 815.0 & 1059.3 & 678.7 & 841.6 \\
		Classfication        & BMS   & HSC   & BMS   & BMS    & BMS   & HSC   \\
		Band gap (eV)        & 2.388 & 1.162 & 2.776 & 0.963  & 2.762 & 1.208 \\
		\hline
	\end{tabular}
}
\end{table*}

For the three structures with relatively high exchange energies, namely mp-35596 (\ce{Fe2NiO4}), mp-1214367 (\ce{BaTbFe4O7}), and mp-39239 (\ce{SrLaMnRuO6}), we also calculated their N$\acute{e}$el temperatures, which are 209.1 K, 39.6 K and 216.5 K, respectively (see in Figure S1). Further analysis through partial substitution of homologous elements shows that certain substitutions can yield HSC or BMS, enriching the possibilities for spintronics materials. Detailed substitution results and corresponding structural data are given in Supplementary Information Tables S1.

Finally, all screened ferrimagnetic semiconductors possess negative formation energies based on the SCAN functional (See in Table \ref{Table:1}). For instance, the formation energy ($E_{\mathrm{f}}$) of \ce{Gd3FeO6}(mp-1189831) is the lowest at -3.34 eV/atom, while that of \ce{Li2Fe2(CO3)3}(mp-1290535) and \ce{MnFeO3}(mp-674482) is approximately -1.84 eV/atom. The energy above hull ($E_{\mathrm{abh}}$) value for all materials is less than 0.1 eV/atom, suggesting a likelihood of experimental synthesis. Excitingly, \ce{Mn3O4}(mp-ID: mp-18759, ICSD: 167411) has been synthesized experimentally.

\section*{Conclusion}

In conclusion, we use a systematic high-throughput screening approach to discover 10 ferrimagnetic BMS and 9 ferrimagnetic HSC. Detailed analysis of the electronic and magnetic properties of the identified materials confirmed their potential to generate 100\% spin-polarized carriers, highlighting their suitability for advanced spintronics applications. The successful prediction of several ferrimagnetic semiconductors with high magnetic transition temperatures underscores the effectiveness of high-throughput computational screening for the rapid discovery of novel spintronics materials. In all, our findings provide a valuable resource for the further exploration of intrinsic ferrimagnetic semiconductors and offer a robust framework for the development of spintronics. 

\section*{Methods}

The simulations of electronic structures based on first-principles are performed using the Vienna ab initio Simulation Package (VASP).\cite{Kresse1996,KRESSE199615} These simulations employ the generalized gradient approximation (GGA) of the Perdew, Burke, and Ernzerhof (PBE) exchange-correlation functional,\cite{PhysRevB.50.17953, perdew1996generalized} with collinear spin polarization. For computational consistency, the cutoff for the plane wave basis set, applied to valence electrons, is set at 520 eV. The convergence criteria for the electronic self-consistent field (SCF) iterations and the ionic step iterations are defined as $1.0\times10^{-6}$ eV and $0.5\times10^{-3}$ eV$\AA^{-1}$, respectively. The reciprocal space grid is determined by employing the Pymatgen method, which ensures a grid density of 8000.\cite{Ong2013b} For accurate consideration of onsite Coulomb repulsion among \textit{d} or \textit{f} electrons, we implement the GGA+U scheme\cite{PhysRevB.52.R5467} in the screening phase. Here, the effective onsite Coulomb interaction parameter ($U$) and the exchange interaction parameter ($J$) for various structures are set according to the default values in Pymatgen.\cite{Ong2013b} To address the semi-local PBE calculations for electronic structures, the density of states of the final candidate materials is computed using the screened hybrid HSE06 functional.\cite{heyd2003hybrid, ge2006erratum} This functional incorporates 20\% Hartree-Fock exchange. The crystal structure figure is obtained via VESTA program.\cite{Momma2008}


\section*{Acknowledgements}

\begin{description}
	\item[Author Contributions]  Haidi Wang, Xiaofeng Liu, and XingXing Li conceived the idea. Haidi Wang, Qingqing Feng, and Shuo Li performed the calculations. All authors helped to write and modify the manuscript. 
	\item[Funding] This research was funded by the National Natural Science Foundation of China (22203026, 22203025, 22103020, 12174080, 22273092, 22322304, and 22403024), the Fundamental Research Funds for the Central Universities (JZ2024HGTB0162 and JZ2023HGTB0219), the Anhui Provincial Natural Science Foundation (2308085QB52), the Strategic Priority Research Program of the Chinese Academy of Sciences (XDB0450101), and the USTC Tang Scholar Program. Computational resources were provided by the HPC Platform of Hefei University of Technology and the Supercomputing Center of the University of Science and Technology of China. 
	\item [Code Availability] The maptool can be found at (https://github.com/haidi-ustc/maptool).
\end{description}

\section*{Conflict of Interest}

The authors declare no competing financial interest.

\section*{Data Availability Statement}

The authors confirm that the data supporting the findings of this study are available within the article and its supplementary materials. All of the structures and related information can be downloaded from material project according to the material ID.

\begin{addendum}
\item[References]

\end{addendum}

\footnotesize{
\bibliography{refs}
}

\vspace{3ex}

\end{document}